\documentclass[%
reprint,
nofootinbib,
aps,
]{revtex4-2}

\usepackage{amsmath,amssymb,amsbsy,amstext,amsthm,simplewick,amsfonts,tensor} 
\usepackage{physics} 
\usepackage{graphicx} 
\usepackage{dcolumn} 
\usepackage{bm} 
\usepackage{comment}
\usepackage{enumitem}
\usepackage{hyperref}
\hypersetup{
	colorlinks = true,
	linkcolor = blue,
	citecolor = blue,
	urlcolor = blue
}

\def\l{\left}
\def\r{\right}
\def\({\l(}
\def\){\r)}
\def\[{\l[}
\def\]{\r]}
\def\pd{\partial}

\def\b{\boldsymbol}

\def\cal{\mathcal}

\def\rm{\mathrm}

\newcommand{\ud}{\mathrm{d}}

\begin{document}

\title{A singularity problem for interacting massive vectors}

\author{Zong-Gang Mou}
\email{zgmou@outlook.com}
\author{Hong-Yi Zhang}%
\email{hongyi@rice.edu}
\affiliation{Department of Physics and Astronomy, Rice University, Houston, Texas 77005, USA}%

\date{\today}

\begin{abstract}
Interacting massive spin-1 fields have been widely used in cosmology and particle physics. We obtain a new condition on the validity of the classical limit of these theories related to the non-trivial constraints that exist for vector field components. A violation of this consistency condition causes a singularity in the time derivative of the auxiliary component and could impact, for example, the field's cosmic history and superradiance around black holes. Such a condition is expected to exist generically in many other non-trivially constrained systems.
\end{abstract}

\maketitle


\emph{Introduction.--}
The lack of direct evidence for Weakly Interacting Massive Particles (WIMP) has driven people to explore different dark matter candidates, among which light massive spin-1 (Proca) fields, the so-called ``dark photons'' or vector dark matter, have been drawing more attention in recent years \cite{ParticleDataGroup:2020ssz, Caputo:2021eaa, Antypas:2022asj}. In general Proca fields could have a variety of non-gravitational interactions and thus very rich dynamics. For example, the coupling to an axion field may allow for a significant energy transfer from axions to dark photons, and make the latter the dominant component of dark matter in the present-day universe \cite{Co:2018lka, Agrawal:2018vin}. If the Proca field possesses a nonlinear self-interacting potential or a non-minimal coupling to gravity, it may drive the cosmic inflation in the early universe \cite{Ford:1989me, Golovnev:2008cf}, and support coherently oscillating localized solitonic field configurations called vector oscillons \cite{Zhang:2021xxa}. The existence of strong self-interactions would also weaken the superradiance bounds on ultralight vectors \cite{Baryakhtar:2017ngi, Fukuda:2019ewf}. Moreover, the theory of vector Galileons, whose effective action contains self-interactions with higher-order derivatives, has been constructed by requiring that the equation of motion has 2nd-order time derivatives and yields three healthy propagating degrees of freedom \cite{Heisenberg:2014rta}. As an IR modification of gravity, it has been shown that these self-interacting Proca fields can lead to a viable cosmic expansion history and even alleviate the Hubble tension without sabotaging the success of General Relativity on scales of the solar system \cite{DeFelice:2016yws, deFelice:2017paw, Heisenberg:2020xak}.

Given these manifold applications, it is necessary to examine the consistency of the interacting Proca theory carefully. One guiding principle that often comes into play is the validity of an effective description of the interaction, which may arise from a low-energy approximation of coupling to other fields or non-minimal coupling to gravity. Another standard lore is that theories with ghosts or energies unbounded from below are usually unstable and problematic \cite{Woodard:2015zca, Cline:2003gs, Sawicki:2012pz} and the initial conditions must be restricted in ``islands of stability'' if possible \cite{Pagani:1987ue, Smilga:2004cy}, although there may be some exceptions \cite{Deffayet:2021nnt}. Regarding massive vectors, it is pointed out that if they are non-minimally coupled to gravity, the longitudinal mode may exhibit ghost instabilities and one can not discuss the vector field dynamics in a healthy way \cite{Nakayama:2019rhg, Kolb:2020fwh}. In practice, one performs as many sanity checks as possible to determine the scope of application for the theory in hand. 

In this letter, we will discuss another type of bound that can arise in the \emph{classical} limit of the theory by demanding the absence of a singularity problem for $\dot A_0$, due to the fact that the interacting Proca theory is a non-trivially constrained system, where the auxiliary component $A_0$ can not always be uniquely solved in terms of the canonical fields. A similar type of constraint is also expected for interacting massive spin-2 fields.

In what follows, we will first clarify three consistency conditions and introduce the singularity problem, then a specific example is carried out in detail both analytically and numerically. Implications of our results are discussed finally. We adopt the mostly plus convention for the metric.

\emph{The singularity problem.--}
For definiteness, let us consider a real-valued massive vector field $A_\mu=(A_0, \b A)$ with the Lagrangian\footnote{The analysis that will be carried out may be straightforwardly generalized for complex fields and for other interactions, see final discussions for more details.}
\begin{align}
	\label{Lagrangian}
	\cal L = -\frac{1}{4}F_{\mu\nu} F^{\mu\nu} - V(A_\mu A^\mu) ~,
\end{align}
where $F_{\mu\nu} = \pd_\mu A_\nu - \pd_\nu A_\mu$ and there is no gauge invariance thanks to the potential $V$, which includes a mass term along with self-interactions. A concrete example is the Abelian-Higgs model, where a quartic self-interaction is induced by Higgs exchange in the low-energy limit. By varying the action $S = \int d^4x \cal L$ with respect to the field $A_\nu$, we find the Euler-Lagrange equation $\pd_\mu F^{\mu\nu} - 2V'(A_\mu A^\mu) A^\nu = 0$. In vector notation, it becomes
\begin{align}
	\label{EOM_A0}
	\nabla\cdot \b \Pi + 2V' A_0 = 0 ~,\\
	\label{EOM_Ai}
	\dot{\b \Pi} + \nabla\cp\nabla\cp\b A + 2V'\b A = 0 ~,
\end{align}
where the prime denotes the derivative of the potential in terms of $A_\mu A^\mu$ and we have defined the conjugate field $\Pi_\mu \equiv \pd\cal L/\pd\dot A^\mu = F_{0\mu}$, so
\begin{align}
	\label{EOM_Pi}
	\dot{\b A} = \b\Pi + \nabla A_0 ~.
\end{align}
One more useful equation can be obtained by noting that $F^{\mu\nu}$ is antisymmetric, so that $\pd_\mu (V'A^\mu)=0$. That is
\begin{align}
	\label{EOM_extra}
	-(V' - 2V'' A_0^2)\dot A_0 - 2 A_0 V'' (\b A\cdot \dot{\b A}) + \nabla\cdot ( V' \b A) = 0 ~.
\end{align}
In the language of Hamiltonian mechanics, $\Pi_0=0$ is a primary constraint, equation \eqref{EOM_A0} is a secondary constraint obtained by requiring $\dot \Pi_0 = \delta H/\delta A^0=0$, and equation \eqref{EOM_extra} is a tertiary constraint obtained by requiring the secondary constraint to be preserved in time \cite{Weinberg:1995mt}. The foundation for these derivations is the stationary action principle, in which we have implicitly assumed that the field $A_\mu$ is continuous otherwise the infinitesimal variation $\delta A_\mu$ is ill-defined. By applying the above formalism, therefore, we require a \emph{consistent} classical system to satisfy at least three conditions everywhere:
\begin{enumerate}[label=(\roman*)]
    \item \label{consistency1} The field $A_\mu(t,\b x)$ is real-valued;
    \item \label{consistency2} The field $A_\mu(t,\b x)$ is continuous;
    \item \label{consistency3} The second-class constraints, e.g. \eqref{EOM_A0} and \eqref{EOM_extra}, are respected.
\end{enumerate}

These conditions are not trivial, and we may gain some insights about them by using equations \eqref{EOM_Ai}-\eqref{EOM_extra} and numerically evolving $\b\Pi$, $\b A$ and $A_0$. Given appropriate initial conditions, suppose that $V'-2V''A_0^2$ never becomes 0, then the infinitesimal variations $\delta\b\Pi$, $\delta\b A$ and $\delta A_0$ are always well defined in a infinitesimal time interval $\delta t$, and the field $A_\mu$ will remain smooth and unique all the time. This is indeed the case for free massive fields, where $V(A_\mu A^\mu) = m^2 A_\mu A^\mu/2$. For theories with self-interactions, however, a singularity is encountered if $V'-2V''A_0^2$ becomes 0 at some spacetime point unless $- 2 A_0 V'' (\b A\cdot \dot{\b A}) + \nabla\cdot ( V' \b A)$ also vanishes in an appropriate way to ensure a finite $\dot A_0$, which otherwise causes a discontinuity in $A_0$ and violate at least one of the consistency conditions. Thus maintaining the continuity of $A_0$ at this point needs an over-constraint and requires fine tuning of initial conditions. It is seen that any plausible interacting Proca theories should ensure that such a problem is avoided in its validity, to wit, the field value should never cross the \emph{boundary} in field space $\{\abs{A_0}, \abs{\b A}\}$ specified by 
\begin{align}
	\label{boundary}
	V'-2V'' A_0^2=0 ~.
\end{align}
One may think that the problem identified here can be easily avoided if we use equation \eqref{EOM_A0} instead of \eqref{EOM_extra} to obtain $A_0$. As will be shown shortly, this difficulty is actually independent of whether and how we evolve the system numerically.

\emph{A concrete model.--}
In order to understand the significance of this singularity bound, it is illuminating to consider the simplest possibility of a self-interaction
\begin{align}
	\label{potential_A4}
	V(A_\mu A^\mu) = \frac{m^2}{2} A_\mu A^\mu + \frac{\lambda}{4}(A_\mu A^\mu)^2 ~,
\end{align}
where $A_0$ can be solved in closed form. We are going to show that if we stick with condition \ref{consistency1} and \ref{consistency3}, then condition \ref{consistency2} will necessarily be violated if the field system hits the boundary \eqref{boundary} during its evolution.

\begin{figure}
	\centering
	\includegraphics[width=0.49\linewidth]{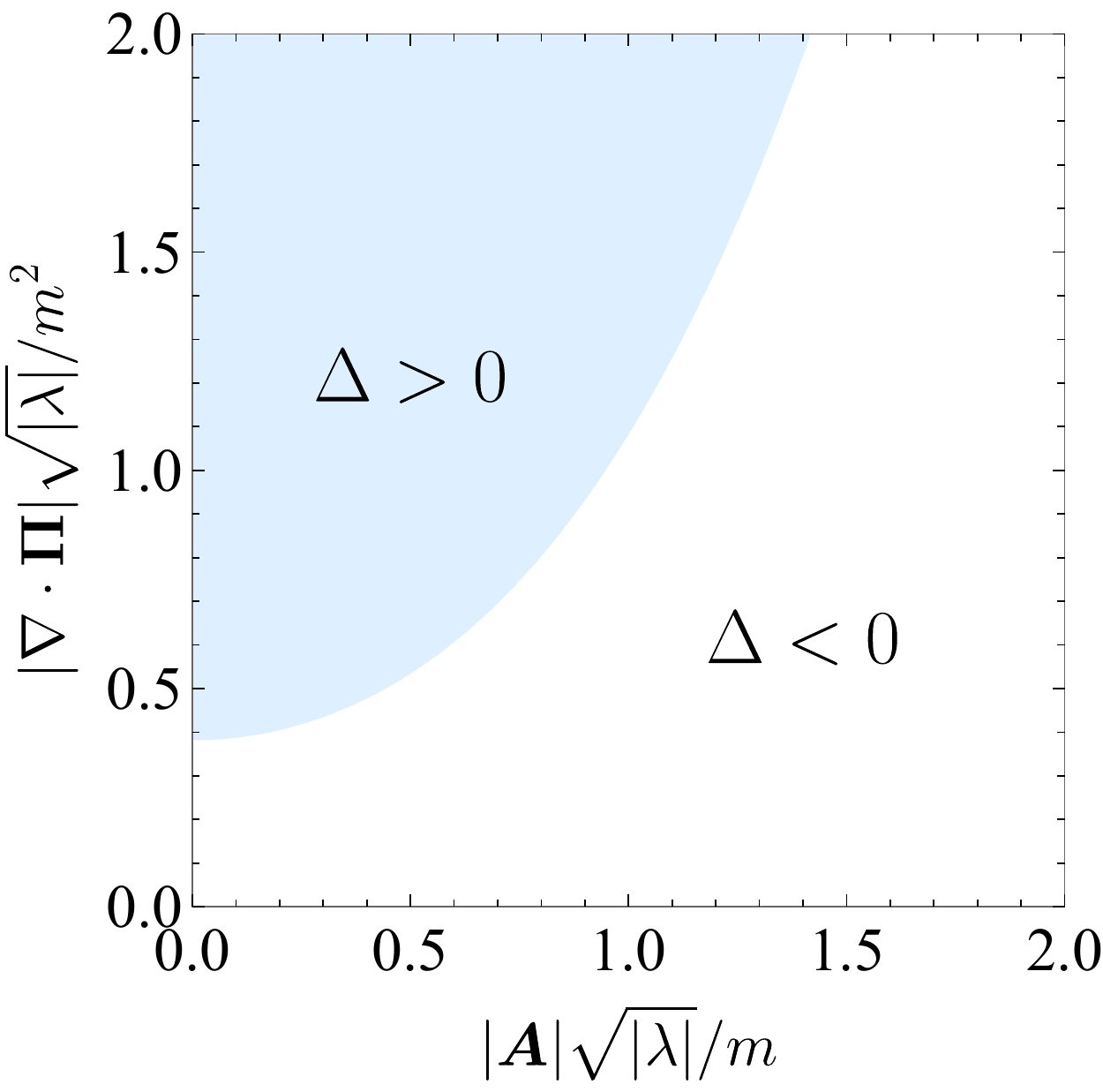}
	\includegraphics[width=0.49\linewidth]{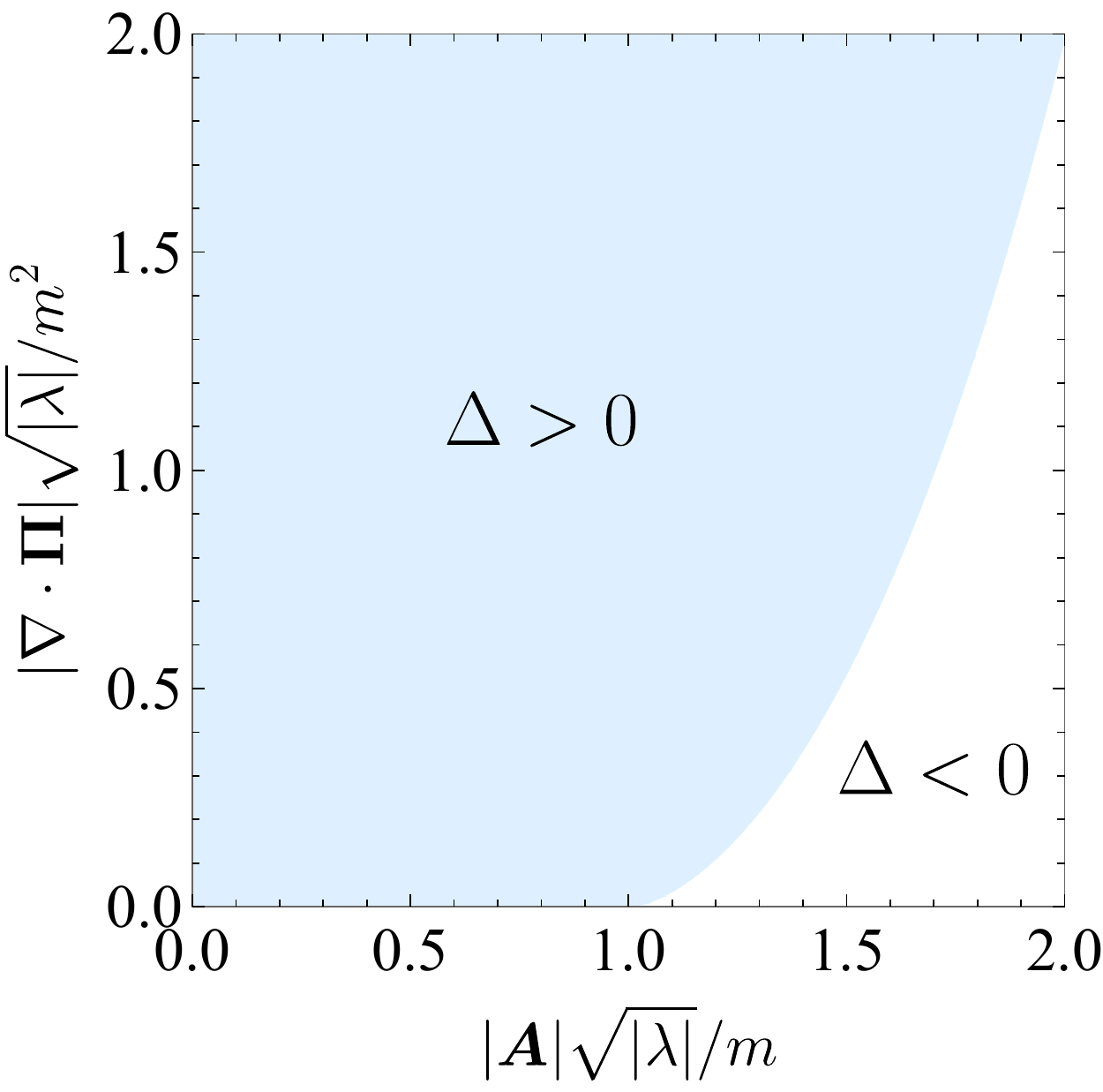}
	\caption{The value of the discriminant $\Delta$ in terms of $\nabla\cdot\b\Pi$ and $\b A$, defined by equation \eqref{Delta}, for repulsive ($\lambda>0$, left) and attractive ($\lambda<0$, right) self-interactions. There are one, two or three different real roots of $A_0$ in equation \eqref{cubic_eq} when $\Delta$ is $>,=$ or $<0$.
	}
	\label{fig:boundary_divPi_Ai}
\end{figure}

The secondary constraint \eqref{EOM_A0} in this case becomes 
\begin{align}
	\label{cubic_eq}
	A_0^3 + c_1 A_0 + c_2=0 ~,
\end{align}
where $c_1 = - m^2/\lambda - \b A^2$ and $c_2 = -\nabla\cdot\b\Pi/\lambda$. The general solution of this cubic equation can be given by the Cardano's formula,
\begin{align}
	\label{root1}
	A_0^{(1)} &= u+v ~,\\
	\label{root2}
	A_0^{(2)} &= \frac{-1+i\sqrt{3}}{2}u + \frac{-1-i\sqrt{3}}{2}v ~,\\
	\label{root3}
	A_0^{(3)} &= \frac{-1-i\sqrt{3}}{2}u + \frac{-1+i\sqrt{3}}{2}v ~,
\end{align}
where $u = \sqrt[3]{- c_2/2 + \sqrt{\Delta}}$ and $v = \sqrt[3]{- c_2/2 - \sqrt{\Delta}}$. The $A_0^{(1)}$ is a real root and the other two are complex conjugate if $\Delta >0$. All three are real roots with $A_0^{(2)}$ and $A_0^{(3)}$ being the same if $\Delta=0$. And all three are different real roots if $\Delta <0$.\footnote{Depending on the sign of $\lambda$, there are three or one real roots when $\lambda\rightarrow 0^+$ or $0^-$. In either case there is only one finite root as expected since the free theory should be recovered in this limit.} Here the discriminant is defined as
\begin{align}
	\label{Delta}
	\Delta\equiv \(\frac{c_2}{2}\)^2 + \(\frac{c_1}{3}\)^3 = \( \frac{\nabla\cdot\b\Pi}{2\lambda} \)^2 - \( \frac{m^2}{3\lambda} + \frac{\b A^2}{3} \)^3 ~.
\end{align}
The value of $\Delta$ in terms of $\nabla\cdot\b\Pi$ and $\b A$ is shown in figure \ref{fig:boundary_divPi_Ai}. 

A few subtleties need to be clarified when we apply the Cardano's formula \eqref{root1}-\eqref{root3}. First, we have defined the square root of any number by its principal value. Second, we have defined the cube root of any number by its \emph{principal} value when $c_2 \le 0$ and its \emph{anti-principal} value when $c_2 >0$. The (anti-)principal cube root returns the real cube root for a real number, and the root with the (smallest) greatest real part for a complex number. These conventions are adopted such that $A_0^{(1)}$ is always real and all three roots are continuous everywhere except at $\nabla\cdot\b\Pi=0$, where $A_0$ can actually remain continuous by switching roots.

If a field system crosses the boundary $\Delta=0$ (and $\nabla\cdot\b\Pi\neq 0$) during its evolution, then the real roots $A_0^{(2,3)}$ will be annihilated or created depending on which region in figure \ref{fig:boundary_divPi_Ai} the system is in before the crossing. It is easy to see that the discontinuity of $A_0$ is an inevitable consequence if $A_0$ follows either $A_0^{(2)}$ or $A_0^{(3)}$, and if the system hits $\Delta=0$ from the white region where $\Delta<0$. 

\begin{figure}
	\centering
	\includegraphics[width=0.49\linewidth]{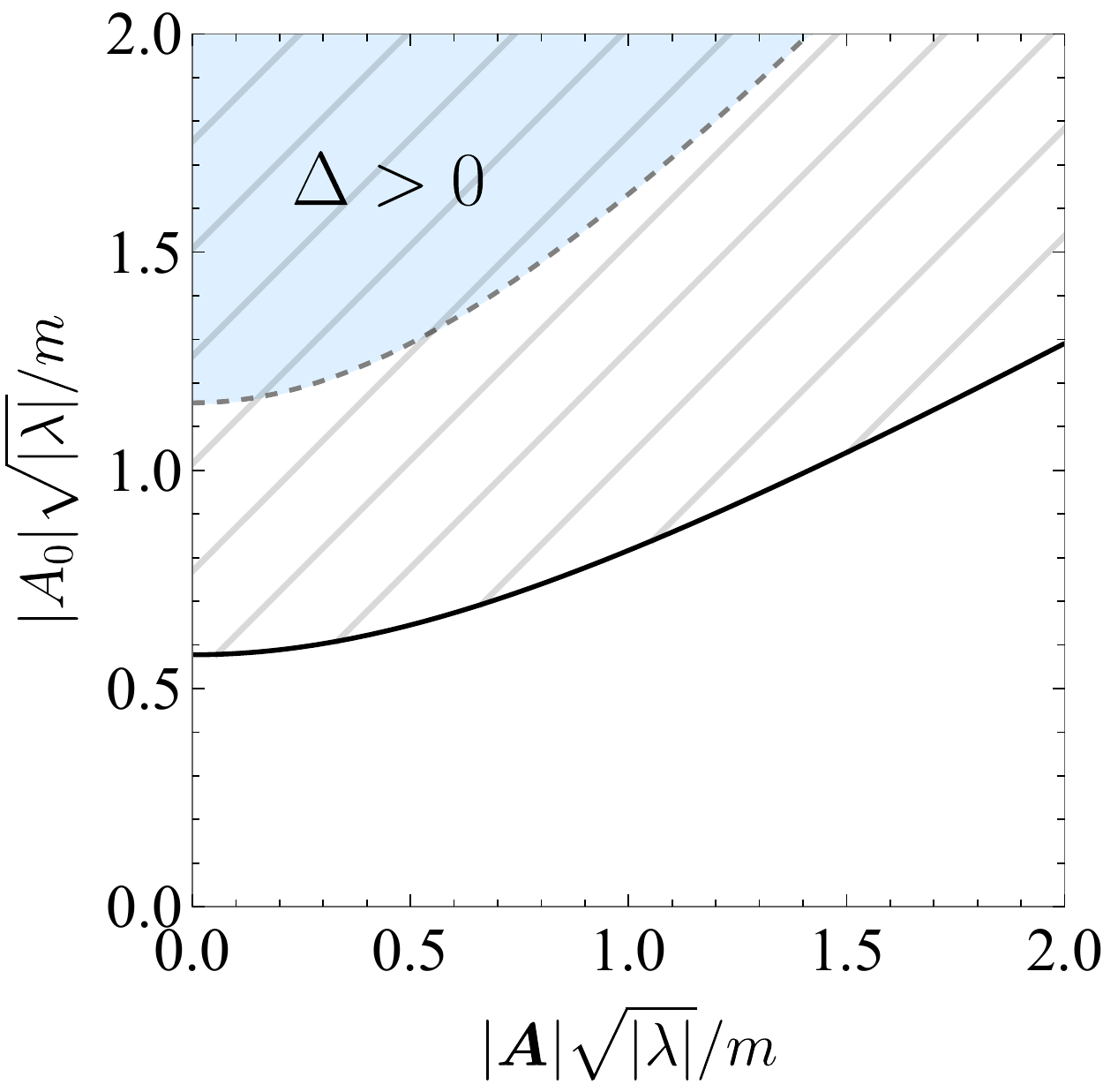}
	\includegraphics[width=0.49\linewidth]{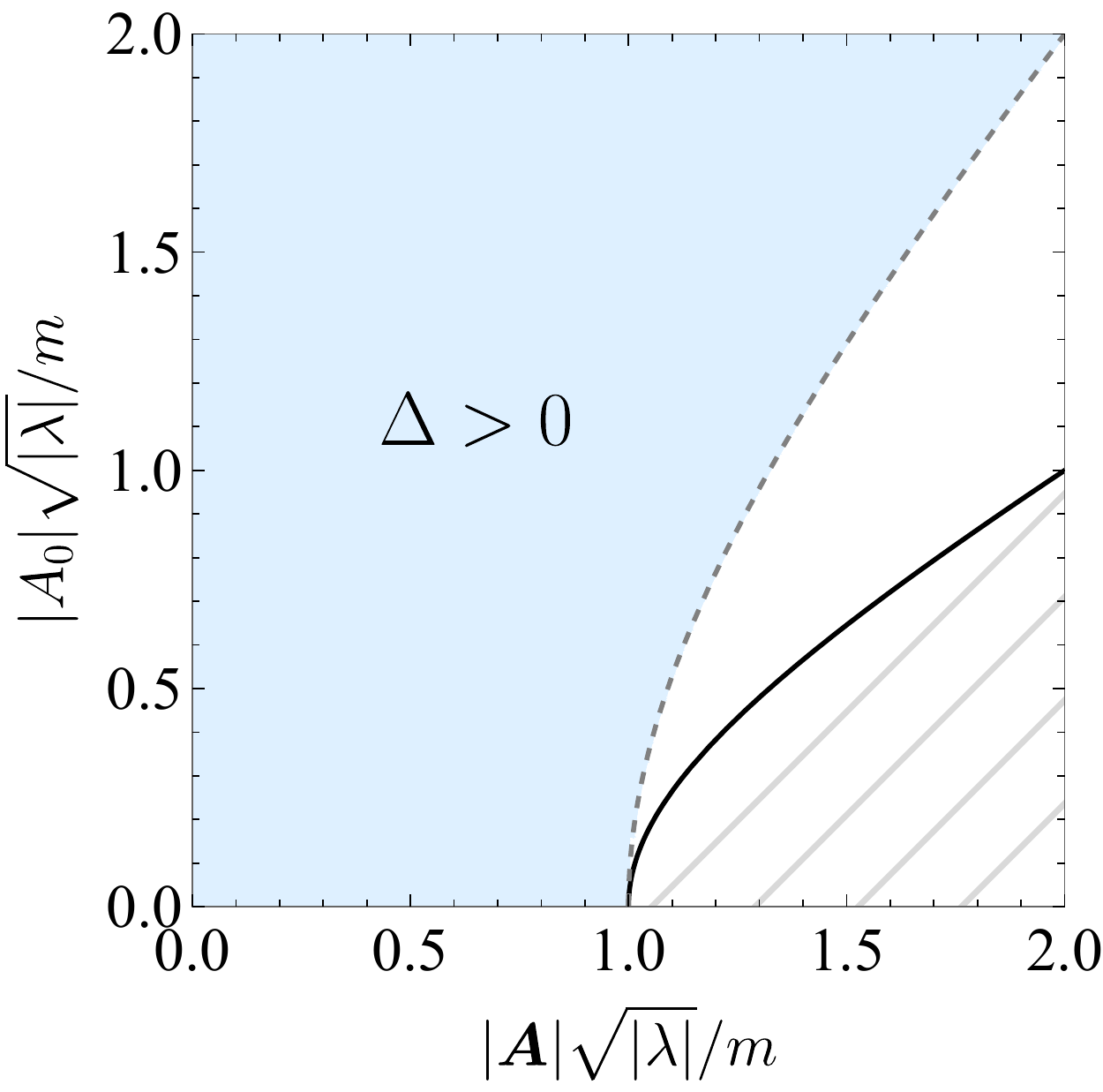}
	\caption{The value of the discriminant $\Delta$ in terms of $A_0$ and $\b A$ for repulsive ($\lambda>0$, left) and attractive ($\lambda<0$, right) self-interactions. The colored and white regions correspond to $\Delta>0$ and $\Delta<0$ as in figure \ref{fig:boundary_divPi_Ai}, and the gray dashed and black solid curves represent $A_0=2 A_{0,\rm{crit}}$ and $A_0=A_{0,\rm{crit}}$ at which $\Delta=0$. A consistent classical system should never cross the black solid curve, which is exactly the one specified by equation \eqref{boundary} (see the texts for proof). Allowing field values to be small, the system during the evolution should never enter into the meshed region.
	}
	\label{fig:boundary_A0_Ai}
\end{figure}

Now we will show that $A_0$ can not remain continuous if the system hits the boundary specified by equation \eqref{boundary}. To do this, we can judiciously rewrite the discriminant $\Delta$ in terms of $|A_0|$ and $|\b A|$ by using the secondary constraint \eqref{EOM_A0}. The value of $\Delta$ in terms of $|A_0|$ and $|\b A|$ is shown in figure \ref{fig:boundary_A0_Ai}. At $\Delta=0$, the three roots \eqref{root1}-\eqref{root3} become $|A_0^{(1)}| = 2A_{0,\rm{crit}}$, $|A_0^{(2,3)}|=A_{0,\rm{crit}}$, where\footnote{We are only interested in the case where $A_{0,\rm{crit}}$ is real.}
\begin{align}
	A_{0,\rm{crit}}(\b A) = \sqrt{\frac{m^2 + \lambda \b A^2}{3\lambda}} ~,
\end{align}
and $A_0=2A_{0,\rm{crit}}$ and $A_0=A_{0,\rm{crit}}$ are visualized as the gray dashed and solid black curves respectively in figure \ref{fig:boundary_A0_Ai}. Note that only $A_0 = A_{0,\rm{crit}}$ (solid black line) corresponds to the boundary $V'-2V''A_0^2=0$, and the adjacent regions separated by this line both have $\Delta<0$, which justifies the foregoing claim.

On the other hand, it is always safe to cross the gray dashed line, since in this case $A_0$ follows the root $A_0^{(1)}$ and $A_0^{(1)}$ is real and continuous. But there is no guarantee that the evolution will be healthy if the root $A_0^{(1)}$ is chosen for $A_0$ initially, because $A_0$ switches roots at $\nabla\cdot\b \Pi = 0$. In order to avoid the singularity problem and also allowing field values to be small, we conclude that the field evolution should be restricted in the non-meshed region in figure \ref{fig:boundary_A0_Ai}.

A minimal model of \eqref{potential_A4} is carried out numerically in $1+1$-dimensional spacetime to support the above analysis. We present field-space trajectories for repulsive self-interactions in figure \ref{fig:numerical1}. The case of attractive self-interactions is similar, and thus only shown in the appendix \ref{sec:numerics}, where numerical details are also provided.

\begin{figure}
	\centering
	\includegraphics[width=0.49\linewidth]{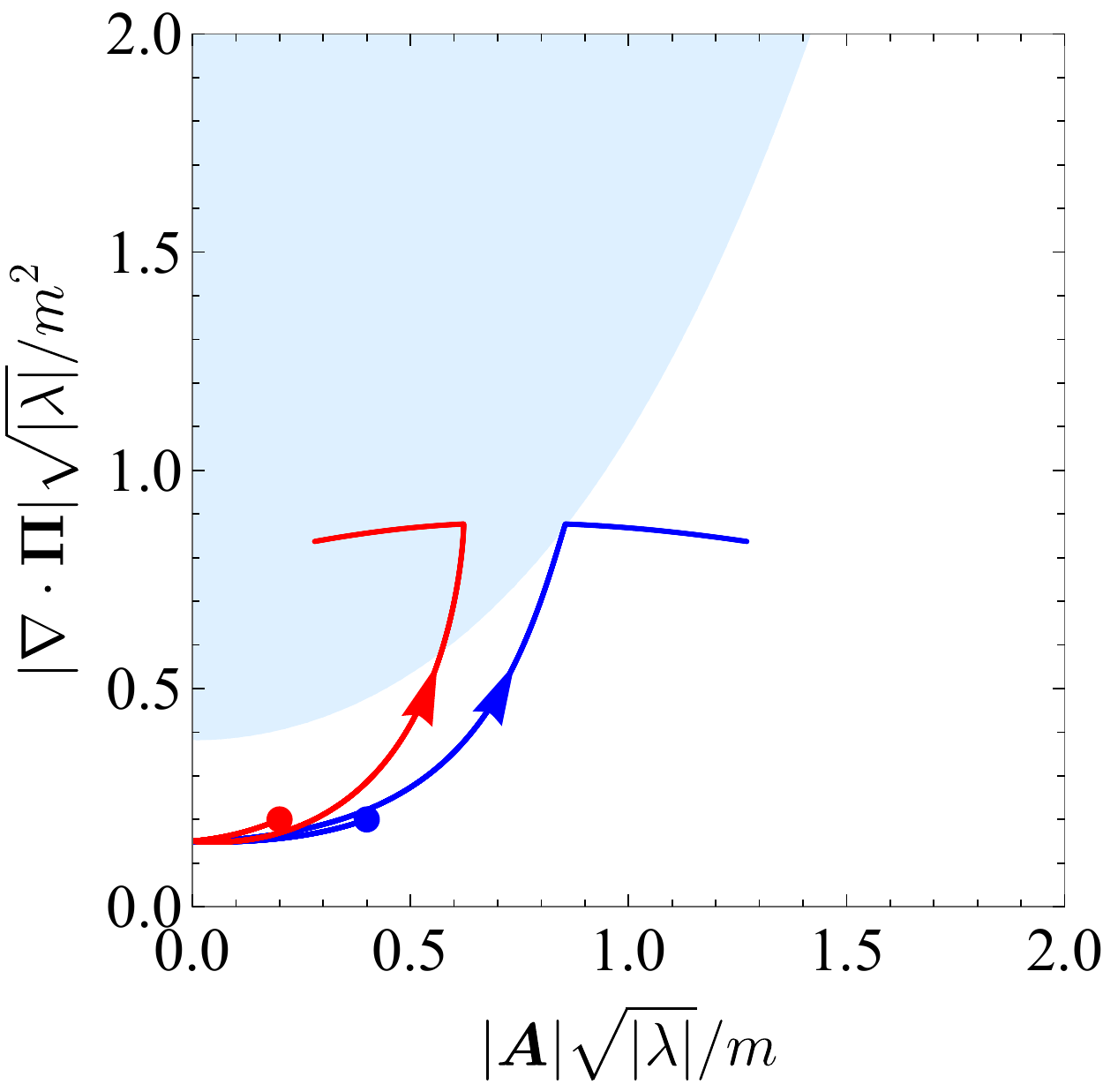}
	\includegraphics[width=0.49\linewidth]{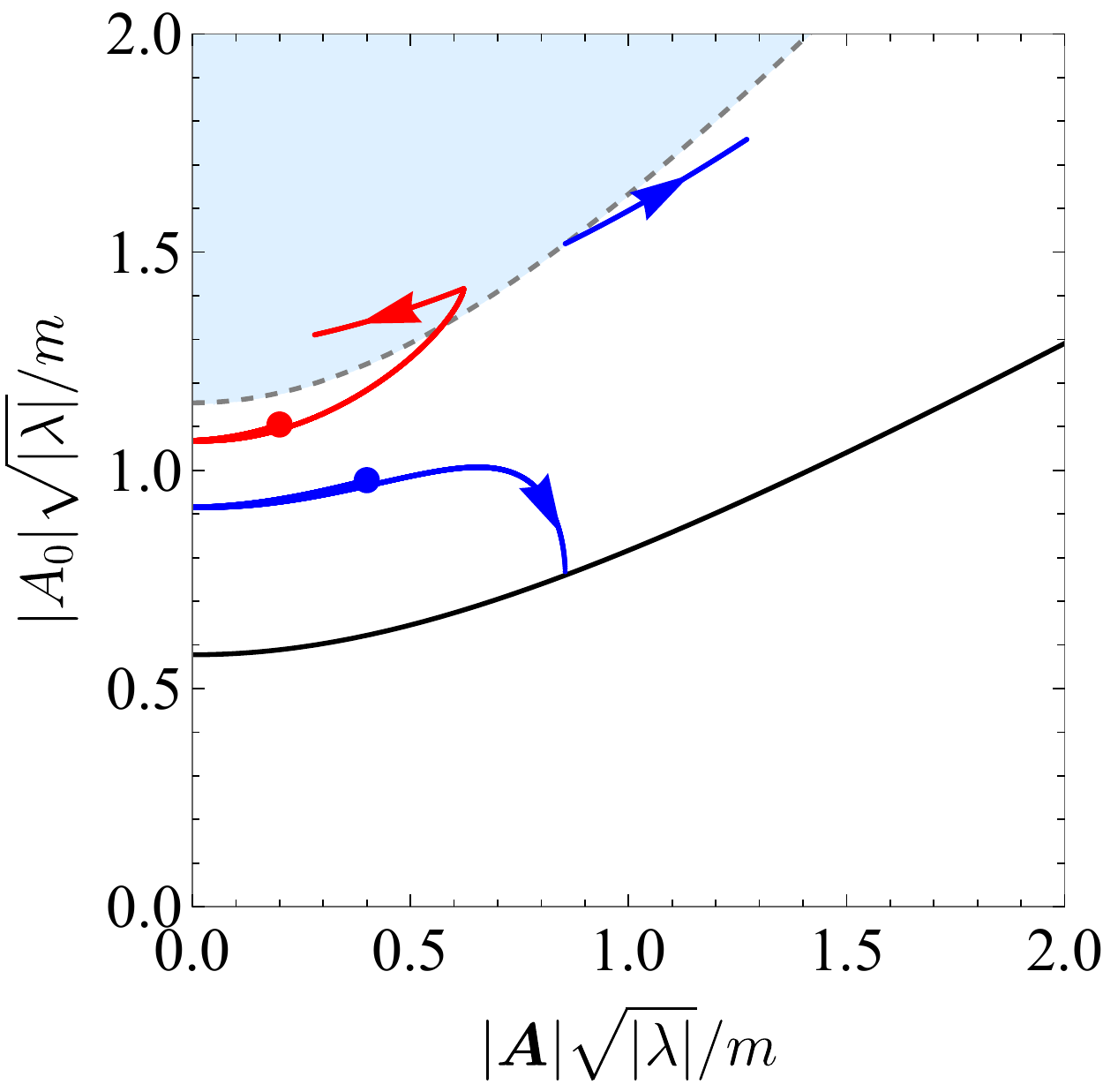}
	\caption{Field-space trajectories of a numerical example for repulsive self-interactions, where the system crosses the black solid boundary specified by equation \eqref{boundary}. The colored and white regions, and the gray dashed and black solid curves have the same meaning as in figure \ref{fig:boundary_divPi_Ai} and \ref{fig:boundary_A0_Ai}. The blue and red trajectories represent the time evolution of fields at two adjacent spatial locations starting from the solid point. As shown by the blue trajectory, when the system meets the black solid boundary, the value of $A_0$ can no longer remain continuous and suddenly jumps to the gray dashed line, which violates the consistency conditions. Numerical details and an example for attractive self-interactions are provided in the appendix \ref{sec:numerics}. 
	}
	\label{fig:numerical1}
\end{figure}

Up to this point, it looks like that the ``discontinuity problem'' is a more appropriate name inasmuch as the temporal component $A_0$ can not be continuous when the field system hits the boundary specified by \eqref{boundary}. In fact, the discontinuity is just an artificial phenomenon because in principle we can stick with the condition \ref{consistency1} and \ref{consistency2} instead, and then we will reach a conclusion that the second-class constraints can not be obeyed. The real problem is that at least one of the consistency conditions will be violated if the boundary \eqref{boundary} is hit, which is closely related to a singularity in $\dot A_0$.

\emph{Conclusions and discussions.--}
We have demonstrated that there exists a generic constraint on field values for interacting Proca theory. Such a constraint can be obtained by observing that a classical massive spin-1 field $A_\mu(t,\b x)$ should be both real-valued and continuous, and the second-class constraints of the system be preserved everywhere. Ensuring these conditions is crucial to get realistic results in some contexts, where non-gravitationally interacting classical fields may play a pivotal role, such as density perturbations in dark photon production \cite{Agrawal:2018vin}, vector oscillons\footnote{Our constraint for the self-interaction considered in \cite{Zhang:2021xxa} implies that vector oscillon fields there are bounded by equation \eqref{boundary} and are unlikely to have very large amplitudes, in constrast to their scalar counterpart with a $\phi^6$ self-interaction \cite{Ibe:2019vyo, Zhang:2020bec}.} \cite{Zhang:2021xxa} and black hole superradiance of vectors \cite{Baryakhtar:2017ngi, Fukuda:2019ewf}. Taking non-derivative self-interactions as a concrete example, we have shown that the field system during its evolution should never meet the singularity bound specified by equation \eqref{boundary}, otherwise the effective description becomes inconsistent.

Loosely speaking, trajectories in phase space (if we could ever visualize them for PDEs) would intersect at the singularity bound \eqref{boundary}, indicating that $A_0$ can no longer be solved uniquely. This situation is usually avoided in physics equations because of the Picard-Lindelof theorem (also called the existence and uniqueness theorem), which states that the existence and uniqueness of solutions are guaranteed if the derivative of the variable is continuously differentiable. To fully appreciate this point, in appendix \ref{sec:toy_model} we provide a toy model with second-class constraints where the phase portrait can be shown explicitly. We also note in there the existence of a basin of attraction towards the singularity bound. If the same goes for interacting massive vectors, the allowed field space is further restricted.

As another example, gauge-invariant interactions (such as those only involving $F_{\mu\nu}$) do not yield the singularity problem, because the gauge-invariant part of the action must satisfy $\pd_\mu (\delta S_\rm{GI}/\delta A_{\mu})=0$ while we have $\pd_\mu (\delta S /\delta A_{\mu})=0$ if the equation of motion is satisfied.\footnote{By gauge invariance here we mean that the action is invariant under the gauge transformation $A_\mu\rightarrow A_\mu + \pd_\mu \Lambda$. More generally, a combined gauge transformation with another scalar field can be invented for massive vectors by using the Stueckelberg trick \cite{Hinterbichler:2011tt}. In this case one must generalize $\pd_\mu (\delta S_\rm{GI}/\delta A_{\mu})=0$ to include the scalar field.} Thus the tertiary constraint like \eqref{EOM_extra} can be obtained solely from gauge symmetry breaking terms. The analysis in this letter may be straightforwardly generalized for more general interactions for Proca fields, and even for interacting massive spin-2 fields.

We note that the singularity bound equally applies for complex fields. To see this, we may separate the real and imaginary part of a complex field $A_\mu = R_\mu + i I_\mu$, then the theory of $A_\mu$ becomes a theory of two interacting real fields $R_\mu$ and $I_\mu$, where each field is constrained by demanding the absence of the singularity.

The existence of the singularity problem in a theory indicates that the theory can not be the complete story. Learning from perturbative unitarity \cite{Lee:1977eg, Schwartz:2014sze}, the standard solution would be to introduce new particles or to look for a UV completion above the scale where the singularity bound is met. For example, we can introduce a Higgs boson to rescue the quartic theory \eqref{potential_A4}. We leave such considerations for future work.

Note: In the final stage of preparation of this work, we became aware of another related forthcoming work by Clough, Helfer, Witek and Berti, which considers ghost instabilities in massive self-interacting vector fields. We will compare it with our work in the future.

\begin{acknowledgments}
We would like to thank Mustafa Amin, Ray Hagimoto, Soichiro Hashiba, Mudit Jain, Siyang Ling and Andrew Long for fruitful discussions. We would especially like to thank Mustafa Amin and Andrew Long for a careful reading of the manuscript and suggestions for its improvement. HYZ is partly supported by DOE-0000250746.
\end{acknowledgments}

\appendix

\section{Numerical details}
\label{sec:numerics}

As proved in the main body of the letter, if we demand that the field is real-valued and satisfies all constraint equations, the value of $A_0$ can not remain continuous when the boundary of the singularity is met. This conclusion can be verified with numerical simulations. 

For simplicity, we may consider a Proca field in 1+1-dimensional spacetime
\begin{align}
	\label{Lagrangian_1d}
	L = \int \ud x \[ \frac{1}{2} F_{01}^2 - V(-A_0^2 + A_1^2) \] ~.
\end{align}
A minimal model after discretization consists of two spatial sites $x_1$ and $x_2$ with periodic boundary conditions. For notation convenience, we define fictitious points $x_3\equiv x_1$ and $x_0\equiv x_2$. Then the task of solving PDEs numerically can be simplified into a problem involving a set of ODEs by defining fields at each site
\begin{align}
	\nonumber
	X_i(t) &\equiv A_0(t, x_i) ~,\\
	\nonumber
	Y_i(t) &\equiv A_1(t, x_i) ~,\\
	\nonumber
	V_i(t) &\equiv V(-X_i^2 + Y_i^2) ~,
\end{align}
where $i=0,1,2,3$ (while those fields with index $0$ and $3$ are fictitious), and the conjugate field $Z_i(t)\equiv F_{01}(t,x_i)$ becomes
\begin{align}
	\label{EOM_Y}
	Z_i=\dot Y_i - \frac{X_{i+1} - X_{i}}{\ud x} ~,
\end{align}
where we have used forward difference to approximate spatial derivatives and $\ud x=x_2-x_1$.\footnote{The numerical system is usually unstable if spatial derivatives are discretized by forward/backward difference method, but the numerical stability is actually irrelevant here since we do not solve the PDEs directly.} As a result, the Lagrangian \eqref{Lagrangian_1d} becomes
\begin{align}
	L = \frac{\ud x}{2} \sum_{i=1}^{2} \[ \frac{1}{2} Z_i^2 - V_i \] ~,
\end{align}
where we have used the trapezoidal rule to approximate the integral, and thus the governing equations corresponding to \eqref{EOM_A0}-\eqref{EOM_extra} are
\begin{align}
	\label{EOM_XYZ}
	&\frac{Z_i - Z_{i-1}}{\ud x} + 2 X_i V_i' = 0 ~,\\
	\label{EOM_Z}
	&\dot Z_i = -2 Y_i V_i' ~,\\
	\label{EOM_X}
	&\dot X_i = \frac{-4 X_i V_i'' Y_i \( Z_i + \frac{X_{i+1} - X_{i}}{\ud x} \) + \frac{2V_i' Y_i - 2 V_{i-1}' Y_{i-1}}{\ud x}}{2 V_i' - 4 V_i'' X_i^2} ~.
\end{align}
Since we may meet a numerical singularity in equation \eqref{EOM_X} if the denominator on the RHS becomes 0, equations \eqref{EOM_Y}-\eqref{EOM_Z} are used to evolve the system. The equation \eqref{EOM_XYZ} may yield one, two or three real roots for $X_i$. Initially we may choose arbitrary one, but during the evolution we always take the closest value to $X_i(t-\ud t)$ as the solution of $X_i(t)$ to ensure its continuity to the most extent. The value of $\ud x$ is not essential to this problem, and we simply set it unity.

An example for repulsive self-interactions is given in figure \ref{fig:numerical1} in the main body of the letter. Initially we set $Y_1=0.4$, $Y_2=0.2$, $Z_1=-0.2$ and $Z_2=-0.4$, and there are three real roots for each $X_i$ since the system stays in the region where $\Delta<0$. The explicit solution shown in figure \ref{fig:numerical1} corresponds to the root $A_0^{(2)}$ for $X_1$ and $A_0^{(1)}$ for $X_2$, i.e. $X_1=-0.977$ and $X_2=-1.105$. The blue and red curve shows the evolution at $x_1$ and $x_2$ respectively. This is consistent with our analytical analysis: only $A_0^{(1)}$ can cross the gray dashed boundary and only $A_0^{(2,3)}$ can cross the black solid boundary.

\begin{figure}
	\centering
	\includegraphics[width=0.49\linewidth]{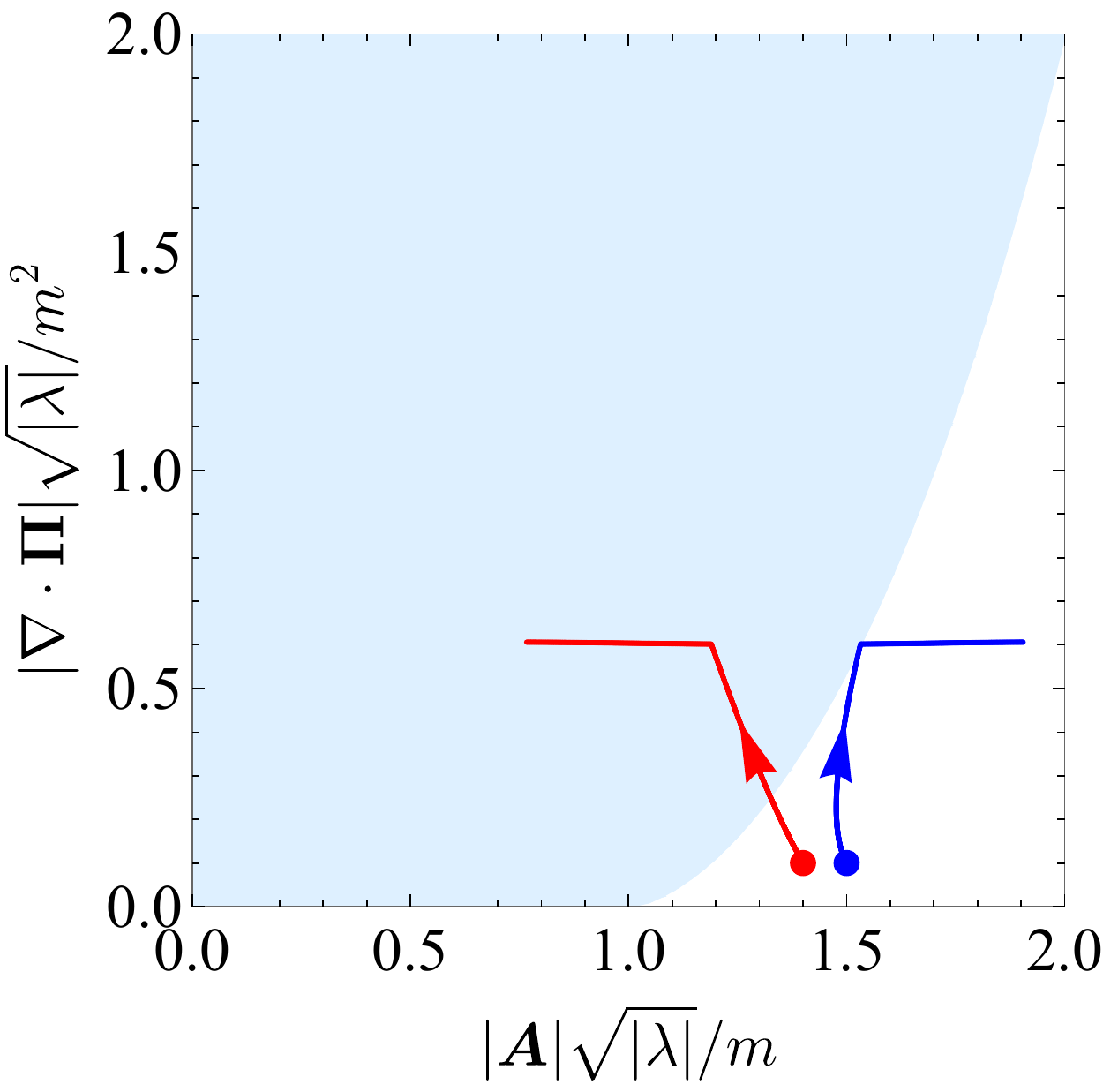}
	\includegraphics[width=0.49\linewidth]{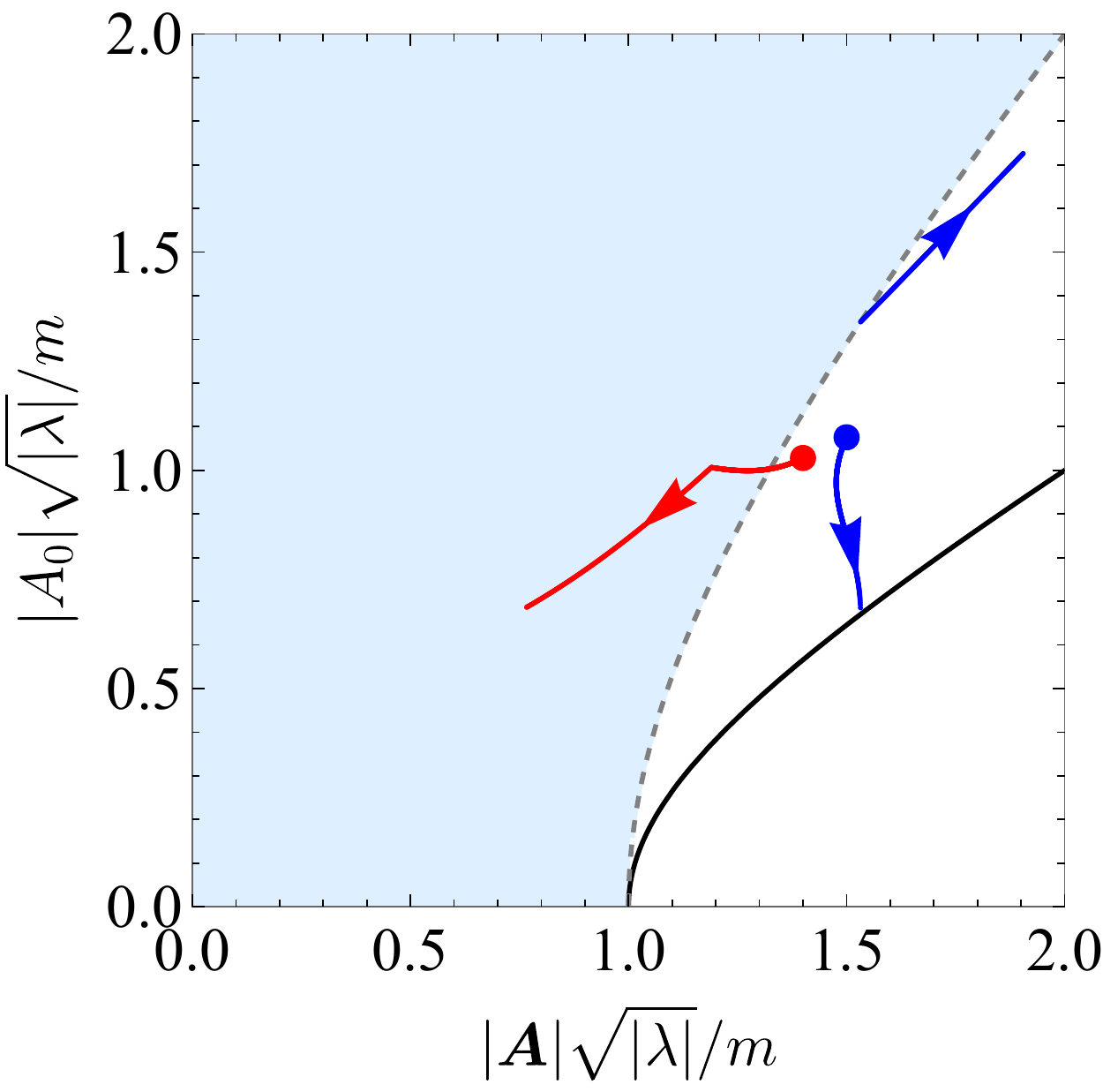}
	\caption{Field-space trajectories of a numerical example for attractive self-interactions, where the system crosses the black solid boundary specified by equation \eqref{boundary}. The notation is the same as in figure \ref{fig:boundary_divPi_Ai}, \ref{fig:boundary_A0_Ai} and \ref{fig:numerical1}.
	}
	\label{fig:numerical23}
\end{figure}

A similar example for attractive self-interactions is presented in figure \ref{fig:numerical23}. The initial conditions are $Y_1=1.5$, $Y_2=1.4$, $Z_1=-0.1$ and $Z_2=-0.2$, and we take root $A_0^{(3)}$ and $A_0^{(1)}$ for $X_1$ and $X_2$ respectively, i.e. $X_1=1.076$ and $X_2=1.028$.

\section{A toy model}
\label{sec:toy_model}

\begin{figure}
	\centering
	\includegraphics[width=0.8\linewidth]{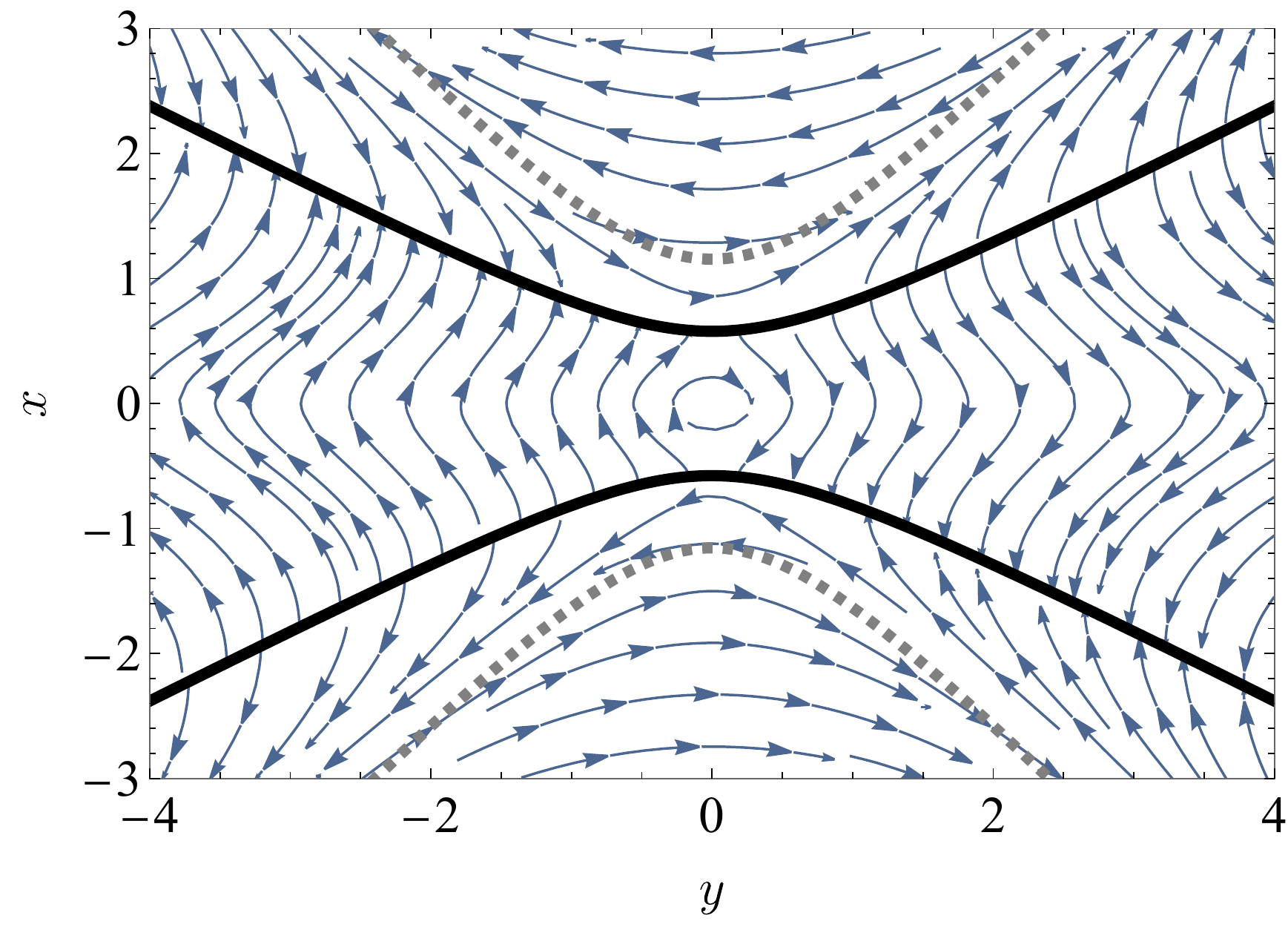}
    \caption{Phase portraits of the toy model. The black solid and gray dashed lines correspond to $\Delta=0$, and the former is also the singularity bound specified by equation \eqref{toy_boundary}. It is seen that there is a basin of attraction towards the singularity bound, where different trajectories intersect.}
    \label{fig:phase_portrait}
\end{figure}

In this appendix, we study the phase space evolution of a toy model that is constrained by second-class constraints just like a Proca system. Consider a Lagrangian
\begin{align}
    L = \frac{1}{2}\left(\dot y-x\right)^2 - V(x,y) ~,
\end{align}
where $x$ and $y$ are real-valued variables, and the potential is
\begin{align}
    V(x,y) = a(-x^2+y^2)+b(-x^2+y^2)^2 ~.
\end{align}
This model is invented such that $x$ and $y$ mimic $A_0$ and $\b A$ in Proca theory, and since it is equipped with ODEs we can draw a complete phase portrait for it. By carrying out the same analysis, it is easy to find the singularity bound
\begin{align}
    \label{toy_boundary}
    V'-2x^2V''=0 ~,
\end{align}
where the prime denotes derivative of the potential with respect to $-x^2+y^2$. This corresponds to the bound \eqref{boundary} for self-interacting vectors. The canonical conjugate variable for this model is $z \equiv \ud L/\ud\dot y= \dot y - x$ for $y$ and 0 for $x$. Thanks to the constraint, $z$ can be solved in terms of $x$ and $y$ and the phase space is actually 2-dimensional, which is shown in figure \ref{fig:phase_portrait}.

Similar to the quartic model in the main text, the discriminant of the model is $\Delta \equiv\frac{1}{4}\left(\frac{z}{4b}\right)^2-\frac{1}{27}\left(\frac{a}{2b}+y^2\right)^3$, where there are one, two or three different real roots for $x$ if $\Delta$ is $>,=,< 0$. In figure \ref{fig:phase_portrait}, both the black solid and gray dashed lines correspond to $\Delta=0$, but only the former is the bound specified by equation \eqref{toy_boundary}. From the trajectories we see that there is actually a \emph{basin of attraction} for the singularity bound, where different trajectories intersect. It is easy to see that finite time is needed to hit the singularity bound if a point in phase space is attracted towards it. The critical exponent behaves like $|x-x_0|\sim |t-t_0|^{1/2}$ as the point approaches the bound. If this is also true for interacting massive vectors, the allowed field space is further restricted.

\bibliographystyle{apsrev4-2}
\bibliography{ref}
\end{document}